\documentclass[final,5p,times,twocolumn]{elsarticle}
\usepackage{amsmath}
\usepackage{amsfonts}
\usepackage{amssymb}
\usepackage{graphicx}
\usepackage{xcolor,soul}
\usepackage{fix-cm}
\usepackage{microtype}
\usepackage{accents}
\usepackage{url}

\journal{}

\graphicspath{{im/}}

\begin{document}

\begin{frontmatter}

\title{Heterarchical Granular Dynamics}

\author[usyd]{Benjy Marks}
\author[usyd]{Shivakumar Athani}
\author[usyd]{Itai Einav}

\affiliation[usyd]{organization={School of Civil Engineering},
            addressline={}, 
            city={The University of Sydney},
            postcode={2006}, 
            state={NSW},
            country={Australia}}



\begin{abstract}
The two most commonly used methods to model the behaviour of granular flows are discrete element and continuum mechanics simulations. These approaches concentrate on the deterministic description of particle or bulk material motion. Unlike these approaches, this paper introduces an alternative model that describes the stochastic dynamics of the void spaces under the action of gravity. The model includes several key phenomena which are observed in deforming granular media, such as segregation, mixing, and an angle of repose. These mechanisms are modelled heterarchically using both spatial and microstructural internal coordinates. Key aspects of the model include its ability to describe both stable and flowing states of granular media based on a solid fraction cut-off, and the influence of particle size on flow, segregation, and mixing. The model is validated with simulations of column collapse and silo discharge.
\end{abstract}




\end{frontmatter}

\section{Introduction}\label{intro}
Granular materials are typically modelled either as a set of discrete particles, or as a continuum. When describing the motion of granular materials, it is generally the dynamics of the particles, rather than the voids, that is of interest to modellers \cite{cundall1979discrete}. This is in contrast to the study of solid materials, where dislocations (the absence of grains) are often studied in and of themselves, e.g.\ \cite{shekhawat2016toughness}. The analogous movement of voids through a granular medium, however, has not attracted the same level of attention, with a few notable exceptions \cite{litwiniszyn1958statistical,mullins1972stochastic,bourdeau1989stochastic,kamrin2007stochastic,kamrin2012nonlocal}.




Here, we introduce a model for the motion of granular material flowing under the action of gravity. The motion of the material is simplified and closely follows the scheme first proposed in \cite{litwiniszyn1958statistical,mullins1972stochastic}, with additional rules to capture the effects of segregation, mixing, and internal stability of particles. In particular, a new model for stability is defined based on the density field~
(\cite{dunatunga_kamrin_2015}, Eq. (2.14)), and is shown to be essential to capture the formation of stable slopes. Furthermore, ideas from stochastic lattice models \cite{marks2011cellular,marks2015mixture,marks2017heterarchical} are used to capture segregation and mixing in terms of particle sizes and voids.

The model proposed here is a heterarchical multiscale model \cite{marks2017heterarchical,bisht2024heterarchical1,bisht2024heterarchical2}. This modelling paradigm involves explicit representation of the microstructure as an internal coordinate of the system. Generally, a representative property of a system, for example the density field, would be described as a function of space and time as $\rho(\vec{x},t)$. To manage information on a scale smaller than the scale at which the density field is defined (which we term the ``representative'' scale), typically a second model is required, with the two models arranged hierarchically, one feeding into the other. Here, however, we introduce a microstructural coordinate which is orthogonal to space and time (an ``internal'' coordinate), so that information at a smaller scale can be described in the one model. We call models that have this property ``heterarchical''.

\begin{table}[t]
    \centering
    \caption{Examples of numerical methods for simulating material behaviour.}
    \begin{tabular}{c|ccc}
         & \textbf{Discrete} & \textbf{Stochastic} & \textbf{Continuum} \\ \hline
        \textbf{Fluid}    &  MD & LBM & CFD \\
        \textbf{Granular} & DEM & --- & FEM \\
        \multicolumn{2}{c}{} & $\uparrow$ & \\
        \multicolumn{2}{c}{} & HGD &    
    \end{tabular}
    \label{tab:methods}
\end{table}

We can define a representative volume element (RVE) by averaging along the internal coordinate. By discretising this internal coordinate, it is possible not only to represent sub-RVE information, but also to allow this information to be exchanged between RVEs. This idea was first proposed in \cite{marks2017heterarchical} with one spatial dimension and one internal coordinate. This work was recently extended \cite{bisht2024heterarchical1,bisht2024heterarchical2} into two spatial dimensions, where the kinematics were separated from the grainsize dynamics (the temporal and spatial evolution of the particle size field) in the context of rotary mills that can be idealised to operate at a steady state, where the kinematics do not change over time. In this work, we extend this paradigm one step further to explicitly include the evolution of the kinematics, which here are driven by the motion of voids. 

The objective of this model is to capture the following behaviours which are characteristic of deforming granular media:

\begin{enumerate}
    \item The material should be stable (i.e.\ not move) when the solid fraction is above a critical value.
    \item The material should be unstable (i.e.\ move) when the solid fraction is below the same critical value.
    \item The motion should be controlled via a single parameter, and the displacement fields produced by varying this parameter should qualitatively represent a variety of granular media.
    \item The motion should stop at a reproducible angle of repose.
    \item The material should segregate by size when moving.
    \item Self-induced diffusion should prohibit perfect segregation (i.e.\ it should induce mixing).
    \item The stress at any point in space should be able to be evaluated. 
\end{enumerate}

We investigate the behaviour of the model by simulating quasi-static granular column collapse \cite{thompson2007granular,degaetano2013influence,lai2017collapse,he2021experimental}. Under such conditions, upon cessation of motion, the angle of repose of the material $\varphi$ can be measured as the angle between the free surface and the horizontal plane. 

It is important to note that the model described here is not the limit of what can be described by integrating the stochastic void migration~\cite{litwiniszyn1958statistical,mullins1972stochastic} and  heterarchical~\cite{marks2017heterarchical} stochastic modelling schemes. Instead, this paper presents a first step towards a general stochastic modelling paradigm for granular media that we term ``Heterarchical Granular Dynamics'' (HGD). As shown in Table~\ref{tab:methods}, we envisage HGD as a general modelling framework for problems related to granular media, in much the same way that the Lattice Boltzmann Model (LBM) is a stochastic modelling paradigm for fluid flow. While LBM sits alongside Molecular Dynamics (MD) and Computational Fluid Dynamics (CFD), we consider HGD to sit between the Discrete Element Method (DEM) and Finite Element Method (FEM) for granular media. While this initial approach omits significant physics, such as inertia and particle-fluid interactions, these and other effects can be added in future implementations. For example, we show in Section~\ref{sec:stress} how stochastic stress propagation models (e.g.~\cite{harr1977mechanics}) can be incorporated as a first step to resolving inertial effects.

\section{Heterarchical Granular Dynamics}

\begin{figure}
    \centering
    \includegraphics{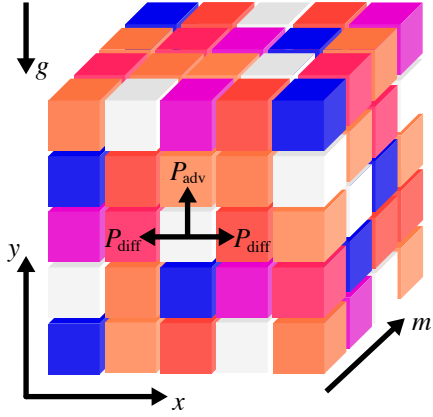}
    \caption{Schematic diagram of the Heterarchical Granular Dynamics model. Each cell in the lattice carries a volume of grains of a particular size (here indicated by the colour, see below Figures for the colour scale). The voids, shown in white, have four degrees of freedom with probabilities of moving left, right, upwards or staying at the same location, as indicated by arrows. Gravity points in the $-y$ direction}
    \label{fig:schematic}
\end{figure}

The model is defined on a heterarchical cartesian grid of dimension 3. The first two dimensions ($x$ and $y$) represent physical space. The third dimension ($m$) represents a microstructural coordinate~\cite{marks2017heterarchical}. If so desired, an additional spatial dimension ($z$) could be added simply. The indices of each cell along these dimensions are labelled ($i,j,k$), corresponding respectively to ($x,y,m$). The number of cells in each respective direction is labelled $X$, $Y$ and $M$. We encode the information of the grain size distribution and the solid fraction at a location $(x,y)$ along the microstructural coordinate with a sufficiently large number of cells $M$ so that the entire coordinate represents a single representative volume element.

The grid is in this way composed of cells, where each cell represents the same volume and contains either:
\begin{enumerate}
    \item Grains of a single size, or
    \item Void space.
\end{enumerate}

The size of the particles in a cell ($i,j,k$) is denoted by $s_{i,j,k}$. If the cell is a void, we set $s$ to zero. We define the number of voids at a location ($i,j$) as

\begin{equation}\label{eq:N}
  N_{i,j}=\underset{\forall s_{i,j,k}=0}{\sum_{k=1}^{M}}1.
\end{equation}

\subsection{Representative measures}

Many properties of a granular medium exist only at the RVE scale, and are undefined for individual cells. We can calculate the porosity $n_{i,j}$ (the volume of voids divided by the total volume) of a representative volume by dividing the number of voids there by the total number of cells in the microstructural direction as

\begin{equation}
n = \frac{N}{M},
\end{equation}

\noindent where we have dropped the indices $i,j$ for brevity, as done for the remainder of the text where appropriate. The solid fraction is defined as $\nu\equiv 1-n$. We also define two alternative measures of the mean local size, the arithmetic and harmonic means, respectively

\begin{align}
    \bar{s} &= \frac{1}{\nu M}\underset{\forall s_k\neq0}{\sum_{k=1}^{M}}s_k,\\
    \overline{s^{-1}} &= \left({\frac{1}{\nu M}\underset{\forall s_k\neq0}{\sum_{k=1}^{M}}\frac{1}{s_k}}\right)^{-1}.\label{eq:means}
\end{align}

\subsection{Void migration}

Voids migrate through the system by swapping with their neighbouring cells. As here we only consider cases where gravity is aligned in the $-y$ direction, the voids advect vertically upward (in the $+y$ direction, as indicated in Figure \ref{fig:schematic}). Additionally, voids diffuse via random motion in the $\pm x$ direction.

We assume that the interstitial vertical velocity is gravity driven, such that without considering inertia particles freefall from one cell to the next before coming to rest. 
The typical velocity can be expressed as

\begin{equation}\label{eq:vertical_velocity}
    u = \sqrt{g\Delta y},
\end{equation}

\noindent where $g$ is the acceleration due to gravity and $\Delta y$ is the grid spacing in the $y$ direction. To represent inertia, it is possible to replace the above relationship with a term derived while considering stress (e.g.\ Section~\ref{sec:stress}) and conservation of momentum. 

We note that particles move in the $-y$ direction as voids move in the $+y$ direction. 
This motion is implemented stochastically, such that cells do not move every time step. When particles do move, they travel a distance $\Delta y$ during a time step $\Delta t$. To resolve the correct velocity, the cells should move with a given probability $\overline{P_\mathrm{adv}}$ such that on average they move with velocity $u = \overline{P_\mathrm{adv}}\tfrac{\Delta y}{\Delta t}$. This probability can be expressed as

\begin{equation}\label{eq:adv}
    \overline{P_\mathrm{adv}} = u\frac{\Delta t}{\Delta y}.
\end{equation}

In addition to this vertical advection, voids can diffuse in the $\pm x$ direction by random motion. The diffusivity in the $x$ direction can be expressed as

\begin{equation}
    D = \frac{\Delta x^2}{2 t_\mathrm{swap}}.
\end{equation}

The typical swapping time $t_\mathrm{swap}$ is the numerical time step $\Delta t$ divided by half the probability of swapping $P_\mathrm{diff}$. The `half' is due to $t_\mathrm{swap}$ being the time to move particles both left and right, whereas we only swap left or right. We therefore have 

\begin{equation}\label{eq:diff}
    P_\mathrm{diff} = D \frac{\Delta t}{\Delta x^2}.
\end{equation}

We make the assumption that this diffusivity is proportional to the advection velocity, with a constant of proportionality $\alpha s$, such that

\begin{equation}\label{eq:D}
\alpha = \frac{D}{u s},
\end{equation}

\noindent and $\alpha$ becomes the inverse of the P\'{e}clet number. As the void itself has no size, the relevant size is that of the particles in the cell into which the void will diffuse. The above relationship between advection velocity and diffusivity has been systematically observed in silo discharge experiments \cite{nedderman1979kinematic,choi2005velocity}. We can also express the average probability of diffusion for a location in space as $\overline{P_\mathrm{diff}}=\alpha u\bar{s}\tfrac{\Delta t}{\Delta x^2}$.

In the case where a neighbour (up, left, or right) is also a void, the probability of swapping in that direction is set to zero. We introduce a size dependence for the advection to account for size segregation. The segregation of fine particles migrating downward through a matrix of larger particles is modelled by dictating that voids are more likely to be filled by smaller particles than by larger ones, as

\begin{equation}
    P_\mathrm{adv} = \overline{P_\mathrm{adv}}\cdot\frac{\overline{s^{-1}}}{s}.\label{eq:seg_scaling}
\end{equation}

The two size measures above are taken from the cell into which the void will advect. This choice of the size dependence of advection is purely empirical, and is chosen to match particle percolation observations \cite{cooke1979interparticle}, whilst also giving rise to no additional bulk displacement of the material (i.e.\ the value of $P_\mathrm{adv}$ averaged over the $m$ coordinate at any location is by definition $\overline{P_\mathrm{adv}}$). The strength of this mechanism could be controlled by, for example, altering the size dependency through $\left(\frac{\overline{s^{-1}}}{s}\right)^a$. 
Here we have assumed $a=1$ for simplicity. 



Finally, we choose the time step $\Delta t$ such that $P_\mathrm{adv} + 2P_\mathrm{diff} \le 1$. As described in \ref{app:dispersion}, this method of time stepping produces numerical dispersion for the advection process. The amount of dispersion can be reduced by lowering the time step $\Delta t$.




\subsection{Stability}

\begin{figure}
    \centering
    \includegraphics{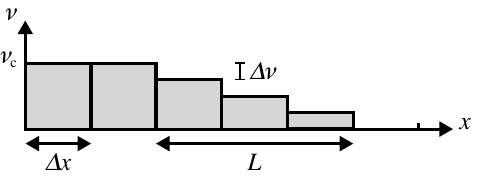}
    \caption{Schematic representation of slope stability criteria.}
    \label{fig:slope}
\end{figure}

We define two states in which an RVE can be: stable or unstable. When unstable, the voids can move according to the rules above. Voids within stable RVEs, however, are not allowed to move out of the cell. Cells can become unstable by voids moving into them from areas that are already unstable. Stability at a location $(i,j)$ is determined by satisfying either of the following two conditions:

\begin{align}
    \nu_{i,j} &\ge \nu_\mathrm{c},\ \mathrm{or}\label{eq:density_cutoff}\\
    \Delta\nu_{i\pm 1,j} &\le \Delta\nu_\mathrm{lim}.\label{eq:slope_stability}
\end{align}


The first condition sets a minimum solid fraction, here termed the \emph{critical solid fraction} $\nu_\mathrm{c}$, below which the material flows and above which it is static, unless the second condition is met.


The second condition defines an angle of repose $\varphi$, which should be equal to the slope of the material at a free surface when the solid fraction of the material is $\nu_\mathrm{c}$. Note that this second condition is only applied near a free surface (within a distance $L$ of a location with $n=0$, see below for the definition of $L$). The threshold value in (\ref{eq:slope_stability}) can be calculated from a geometric argument by considering that the value of $\nu$ should progress from $\nu_\mathrm{c}$ to zero in a distance set by $\mu$.

As depicted in Figure~\ref{fig:slope}, at the free surface a stable slope is defined by an area where the material is at a critical solid fraction $\nu_\mathrm{c}$, and proceeds in $G$ drops of solid fraction $\Delta\nu$, across a distance $L$, to reach $\nu=0$. Under these conditions

\begin{align}
    G &= \frac{\nu_\mathrm{c}}{\Delta \nu}, \\
    L &= (G-1)\Delta x.
\end{align}

We can define the gradient of the slope $\mu=\tan\varphi$ that corresponds to this configuration by noting that across the distance $L$, the height will drop by $\Delta y$, giving

\begin{equation}
    \mu = \frac{\Delta y}{L}
    = \frac{\Delta y}{\Delta x}\frac{1}{\frac{\nu_\mathrm{c}}{\Delta \nu} - 1}.
\end{equation}

Finally, this can be rearranged into an expression for the solid fraction difference required to maintain stability at a free surface as

\begin{equation}
    \Delta\nu_\mathrm{lim} = \frac{\nu_\mathrm{c}}{\frac{\Delta y}{\Delta x}\frac{1}{\mu} + 1}.
\end{equation}

We note that this method of maintaining a stable slope contains inherent discretisation errors, which are discussed in \ref{app:disc}.

\subsection{Stochastic force propagation}

Given the material exists on a regular grid with a well defined density, it is straightforward to implement a force propagation mechanism such as that proposed by Harr in \cite{harr1977mechanics}. This model describes the propagation of vertical forces downwards with a diffusion horizontally due to the heterogeneity of particle contacts. This model has been chosen from several options available in the literature (e.g.~\cite{hrenikoff1941solution,claudin1998models}) due to its relative simplicity, and the ability to predict stress fields which match those observed in granular media. Inspired by \cite{harr1977mechanics}, the propagation of the typical force $F^y$ in the $y$ direction under the effect of self-weight could be represented on a regular grid as

\begin{align}\label{eq:force}
    &F^y_{i, j} = F^y_{i, j+1} + w_{i,j} \nonumber\\
    &+ \gamma \left( F^y_{i+1,j+1} - 2F^y_{i,j+1} + F^y_{i-1, j+1} \right),
\end{align}

\noindent where $w_{i,j}=\nu_{i,j}\rho_s g$ is the weight of the material at $(i,j)$, with $\rho_s$ the solid density, and $\gamma$ controls the horizontal spreading of the vertical forces. Similarly, one can formulate the propagation of the forces $F^x_{i,j}$ along the horizontal $x$ direction. We have not considered in this work the physical relationships between the void migration, force propagation and stability criteria. The above force propagation model could also be improved in the future to ensure momentum conservation in space. Nonetheless, the description of this model serves as a motivation for the structure and scope of HGD models.

\section{Continuum description}

\subsection{The dilute void limit}

As shown previously in other void migration models \cite{litwiniszyn1958statistical,mullins1972stochastic}, it is possible to relate the stochastic rules above to a continuum description of the medium when (1) the voids are dilute (i.e.\ there are so few unstable voids in the system that they do not interact with each other), (2) there are no free surfaces, and (3) the material is everywhere unstable. 
By considering the conservation of voids at a single location $(i,j)$ over a single time step $\Delta t$, the number of voids $N_{i,j}$ at that location updates as

\begin{align}\label{eq:conservation_dilute}
    N_{i,j}^{t+\Delta t} = &N_{i,j}^t - 2\overline{P_\mathrm{diff}}_{i,j}^t N_{i,j}^t \nonumber\\
    +&\overline{P_\mathrm{diff}}_{i+1,j}^t N_{i+1,j}^t + \overline{P_\mathrm{diff}}_{i-1,j}^t N_{i-1,j}^t \nonumber \\
    -&\overline{P_\mathrm{adv}}_{i,j}^t N_{i,j}^t + \overline{P_\mathrm{adv}}_{i,j-1}^t N_{i,j-1}^t.
\end{align}

Rearranging, taking the limit as $\Delta x$, $\Delta y$ and $\Delta t$ approach zero, with porosity $n=\tfrac{N}{M}$, and using Eqs.(\ref{eq:adv}) and (\ref{eq:diff}) we find

\begin{equation}\label{eq:dilute_pde}
    \frac{\partial n}{\partial t} = \frac{\partial^2}{\partial x^2}(\alpha u \bar{s}n) - \frac{\partial}{\partial y}(un).
\end{equation}

This demonstrates that at the dilute void limit, the model represents the advection and diffusion of porosity. We therefore expect the stochastic simulation to converge to the equivalent analytic model by reducing $\Delta x$ and $\Delta t$, and by increasing $M$. We have confirmed this convergence numerically.

\subsection{General void conditions}

In this work, we go well beyond the limit of dilute voids, all the way to $n=1$. It is therefore necessary to extend the previous derivation to include the fact that swapping is not always possible due to the truncated probabilities implied by a void attempting to move into another void. By taking this into account in the upscaling, and neglecting the stability criteria, we recover

\begin{align}\label{eq:conservation_dense}
    N_{i,j}^{t+\Delta t} =N_{i,j}^t -&\overline{P_\mathrm{diff}}_{i,j}^t N_{i,j}^t\left(1 - \frac{N_{i+1,j}^t}{M}\right) \nonumber\\
    -&\overline{P_\mathrm{diff}}_{i,j}^t N_{i,j}^t\left(1 - \frac{N_{i-1,j}^t}{M}\right) \nonumber\\ 
    +&\overline{P_\mathrm{diff}}_{i+1,j}^t N_{i+1,j}^t\left(1 - \frac{N_{i,j}^t}{M}\right) \nonumber\\ 
    +&\overline{P_\mathrm{diff}}_{i+1,j}^t N_{i-1,j}^t\left(1 - \frac{N_{i,j}^t}{M}\right) \nonumber\\ 
    -&\overline{P_\mathrm{adv}}_{i,j}^t N_{i,j}^t \left(1 - \frac{N_{i,j+1}^t}{M}\right) \nonumber\\ 
    +&\overline{P_\mathrm{adv}}_{i,j-1}^t N_{i,j-1}^t\left(1 - \frac{N_{i,j}^t}{M}\right).
\end{align}

Rearranging, and again taking the limit as $\Delta x$, $\Delta y$ and $\Delta t$ approach zero,

\begin{equation}
    \frac{\partial n}{\partial t} = \frac{\partial^2}{\partial x^2}(\alpha u \bar{s}n) - \frac{\partial}{\partial x}(\alpha u \bar{s}n^2)- \frac{\partial}{\partial y}\left(un(1-n)\right).
\end{equation}

If all of the particles are of one size we have that $\frac{\partial\bar s}{\partial x}=0$ and the above equation reduces to

\begin{equation}\label{eq:continuum}
    \frac{\partial n}{\partial t} = \frac{\partial^2}{\partial x^2}(\alpha u \bar{s}n) - \frac{\partial}{\partial y}\left(un(1-n)\right).
\end{equation}

Under these more porous conditions, the behaviour is similar to the dilute case (taking the limit as $n\rightarrow 0$), but the advection in the $y$ direction is reduced in areas where there are many other voids ($n>0$). The diffusion in the $x$ direction is not affected by changes in local porosity.



\begin{figure}
    \centering
    \includegraphics[width=\columnwidth]{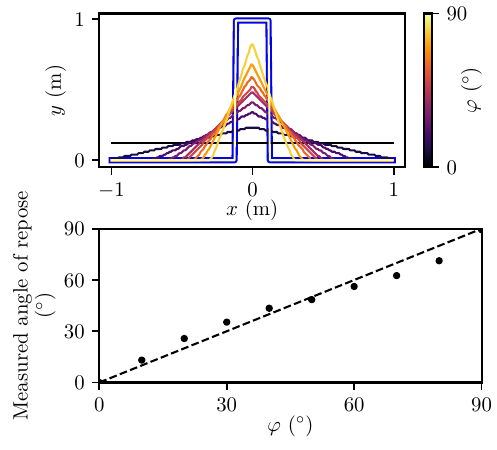}
    \caption{Collapse of monodisperse granular columns initially at the critical solid fraction $\nu_\mathrm{c} = 0.5$. The initial profile is indicated in blue. After collapse, for each of the imposed repose angles $\varphi$ the surface profile from the simulations is obtained (top) and compared against the imposed value (bottom). The dashed line represents equality of the measured and imposed angles of repose.}
    \label{fig:collapse_angle}
\end{figure}

Similar derivations can be made for the motion of the solid phase, see \ref{sec:phi_pde}.





\subsection{Segregation velocity}

As shown in \ref{sec:phi_pde} in Eq. (\ref{eq:rho_hat}), the segregation velocity in the $y$ direction can be expressed as

\begin{equation}\label{eq:seg}
  u_\mathrm{seg} = \left(1 - \frac{\overline{s^{-1}}}{s}\right)un.
\end{equation}

\noindent which recovers the scaling of the segregation velocity proposed in \cite{marks2017size}. The segregation described in Eq. (\ref{eq:seg}) occurs vertically only. We additionally expect segregation to occur in the horizontal direction due to a size-dependent diffusion process. Evidence of these two different mechanisms driving segregation is supported by discrete element simulations in \cite{guillard2016scaling}, see Figure 6 therein, although a quantitative comparison has not been attempted. Validation of this in the future is necessary. For the case of inclined plane flow, which is the most commonly studied geometry for size segregation, we expect that only the advective segregation would be prominent, as the diffusive segregation will occur in the direction of flow and will not be visible. 

\subsection{Stress field}\label{sec:stress}

The force propagation model introduced in Eq.~\ref{eq:force} has a well-defined continuum limit in terms of stress $\sigma$. Under the assumption that $\gamma=\tfrac{\Delta y}{\Delta x^2}\cdot D_\sigma$, this can be shown to be equivalent at the continuum limit to

\begin{equation}
    \frac{\partial\sigma_{yy}}{\partial y} +\rho g= D_\sigma\frac{\partial^2\sigma_{yy}}{\partial x^2},
\end{equation}

\noindent where $D_\sigma$ is the diffusivity of the typical vertical force horizontally, which is assumed to be a function only of elevation~\cite{harr1977mechanics}. Horizontal and shear stresses $\sigma_{xx}$ and $\sigma_{xy}$ can be obtained through $\sigma_{xx}=\tfrac{\partial}{\partial y}(D_\sigma \sigma_{yy})$ and $\sigma_{xy}=-D_\sigma\tfrac{\partial\sigma_{yy}}{\partial x}$. Harr proposes that $D_\sigma\propto h$, with $h$ the depth from the free surface. For full details of the stress model derivation, see~\cite[Chapter 7]{harr1977mechanics}. The first coupling of a void migration deformation model with this stress propagation model was presented in~\cite{bourdeau1989stochastic}. The stress model, its basis in micromechanics and its interaction with the heterarchical coordinate could be improved in future HGD models. These improvements could also serve as the basis for the introduction of inertial effects by revising the probabilities of void migration to conserve momentum.





\section{Results and discussion}

\begin{table}[b]
    \centering
    \caption{Void migration parameters and default values}
    
    \begin{tabular}{ccl}
        Parameter & Value & Description \\ \hline
        $\alpha$ & 4 & Parameter controlling diffusivity \\
        $\nu_\mathrm{c}$ & 0.5 & Critical solid fraction \\
        $\varphi$ & 30$^\circ$ & Angle of repose\\
    \end{tabular}
    \label{tab:parameters}
\end{table}

The original models for the stochastic motion of voids in granular media~\cite{litwiniszyn1958statistical,mullins1972stochastic} were solved analytically in the dilute void limit for a variety of silo discharge scenarios. We do not provide here a quantitative comparison with these previous models, except to note that in the dilute void limit, away from free surfaces, the current model behaves similarly. 
Instead, we validate the new features of the model by examining the case of column collapse. For all examples, unless stated otherwise, the values listed in Table~\ref{tab:parameters} are used in order to describe the void migration, and hence the granular motion. Simulation grid sizes are all $X=100$, $Y=50$ and $M=100$, except for Figure 8 which is $X=50$, $Y=100$ and $M=100$. In order to describe force propagation, we adopt Harr's proportionality with $D_\sigma=h/10$ when comparing with real stress data. 



\begin{figure}[t]
    \centering
    \includegraphics[width=\columnwidth]{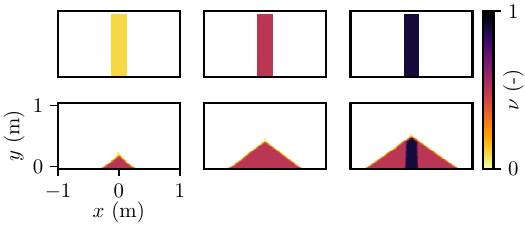}
    \caption{Collapse of monodisperse granular columns with varying initial density. \emph{Left to right}: Solid fractions $\nu=0.1$, 0.5 and 0.9. \emph{Top}: Initial condition. \emph{Bottom}: Final deposit.}
    \label{fig:collapse_fill}
\end{figure}

\subsection{Angle of repose}

When an initially rectangular mass of grains is allowed to collapse under the action of gravity, it eventually forms a static pile. The slope of the free surface of the pile is typically referred to as the \emph{angle of repose}. We are thus able to test our stability criteria, Eqs. (\ref{eq:density_cutoff}) and (\ref{eq:slope_stability}) by allowing a rectangular mass to collapse, and to measure the slope of the free surface arising from a simulation. As shown in Figure \ref{fig:collapse_angle}, the model readily produces static piles at the required angle of repose from 0$^\circ$ to 90$^\circ$.

\subsection{Compaction and dilation}

The model contains a rudimentary mechanism to allow for compaction and dilation in Eq. (\ref{eq:density_cutoff}). Material lower than the critical solid fraction will spontaneously compact towards it, and material that is denser than this limit will need voids to enter to dilate toward it. We can see this effect by observing another case of column collapse, shown in Figure~\ref{fig:collapse_fill}. In the left hand example, material that is initially loose (i.e. the solid fraction is lower than $\nu_\mathrm{c}$) collapses, and forms a static pile at the critical solid fraction. On the right hand side, material that is initially denser than $\nu_\mathrm{c}$ dilates during motion, with the material that is undeformed remaining at its initial density.

There are currently no mechanisms in the model by which a material can be compacted more densely than $\nu_\mathrm{c}$, except by choosing this solid fraction as an initial condition. Given that there are evolving stress and density fields in the model, it is straightforward to add such a mechanism in the future, although this has not been attempted here.

\subsection{Bidisperse mixtures}

\begin{figure}
    \centering
    \includegraphics[width=\columnwidth]{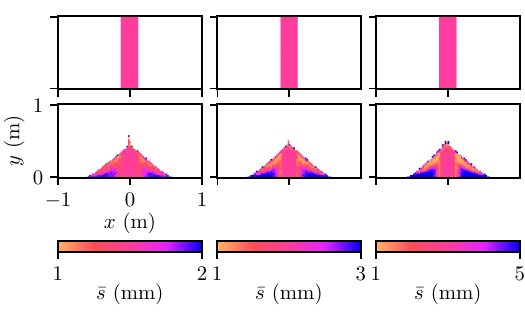}
    \caption{Collapse of bidisperse granular columns with varying size ratio. \emph{Left to right}: Large particles of 2, 3 and 5 mm. Each case has a small particle size of 1 mm, and is composed of 50\% small and large particles. \emph{Top}: Initial condition, where particles are mixed. \emph{Bottom}: Final pile. The colour bar represents arithmetic mean size.}
    \label{fig:collapse_bi}
\end{figure}

A pile of grains of two sizes, when collapsing, segregates by size. As shown in Figure~\ref{fig:collapse_bi}, we find that the larger particles (in blue) migrate in the $x$ direction faster than the finer particles, and so are found below the fine particles at the sides of the pile. With increasing size ratio, we predict an increasingly distinct separation between the zones of small of large particles.

\subsection{Polydisperse mixtures}
\begin{figure}[t]
    \centering
    \includegraphics[width=\columnwidth]{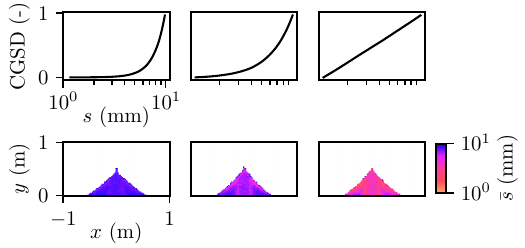}
    \caption{Collapse of polydisperse granular columns with varying grain size distribution. \emph{Left to right}: Particles are drawn from truncated power law distributions with minimum size 1 mm, maximum size 10 mm and a power law exponent of $\beta=-1$, 1 and 3. \emph{Top}: Cumulative grain size distribution. \emph{Bottom}: Final pile. The colour bar represents arithmetic mean size on a logarithmic scale.}
    \label{fig:collapse_poly}
\end{figure}

To demonstrate the ability of the model to predict the segregation patterns from polydisperse materials (i.e. those composed of many particle sizes), we simulate the collapse of piles with truncated power law distributions of the form

\begin{equation}
    F = \frac{s^{3-\beta} - s_m^{3-\beta}}{s_M^{3-\beta} - s_m^{3-\beta}}, 
\end{equation}

\noindent where $F$ is the cumulative grain size distribution, $s_m$ is the minimum particle size, $s_M$ is the maximum particle size and $\beta$ is a power law exponent.

As shown in Figure~\ref{fig:collapse_poly}, polydisperse mixtures segregate in the same manner as bidisperse mixtures, although typically to a lesser degree. There are extremely limited experimental measurements with which to compare the predictions of this model for polydisperse segregation, due to the complexity of measuring particle size distributions experimentally.

\subsection{Stress fields}

\begin{figure}
    \centering
    \includegraphics[width=\columnwidth]{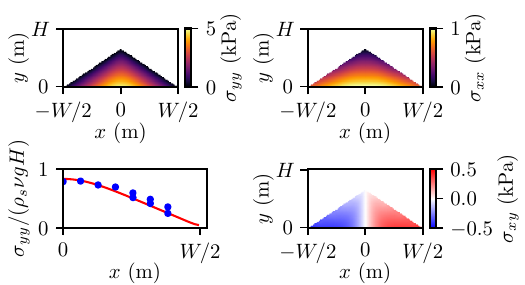}
    \caption{Final state of a monodisperse granular column after collapse of with $\nu_\mathrm{c} = 0.61$ and $\rho_s=2700$~kg/m$^3$, $\varphi=32.5^\circ$ ,$W=1.143$~m, $H=0.36$~m, and $D_\sigma=\tfrac{h}{10}$, where $h$ is the depth from the free surface. Dimensions chosen to match experimental data from~\cite{trollope1956stability}.
    \emph{Top left}: Vertical stress $\sigma_{yy}$. 
    \emph{Top right}: Horizontal stress $\sigma_{xx}$. 
    \emph{Bottom right}: Shear stress $\sigma_{xy}$.
    \emph{Bottom left}: Comparison of basal vertical stress model output (red line) with experimental data from~\cite{trollope1956stability} (blue dots) for a wedge of medium-fine sand with $\varphi=32.5^\circ$ on a rigid base. 
    }
    \label{fig:collapse_stress}
\end{figure}

The stress fields for the case of $\varphi=32.5^\circ$ are shown in Figure~\ref{fig:collapse_stress}. The model predicts increasing normal stresses with depth and peak vertical stress at the centre of the pile. The model results of the vertical stress are compared with experimental data from~\cite{trollope1956stability} for a medium-fine sand deposited onto a rigid plate with $\varphi=32.5^\circ$. The results match favourably given $D_\sigma=h/10$.  

\subsection{Calibration of $\alpha$}

\begin{figure}[t]
    \centering
    \includegraphics[width=\linewidth]{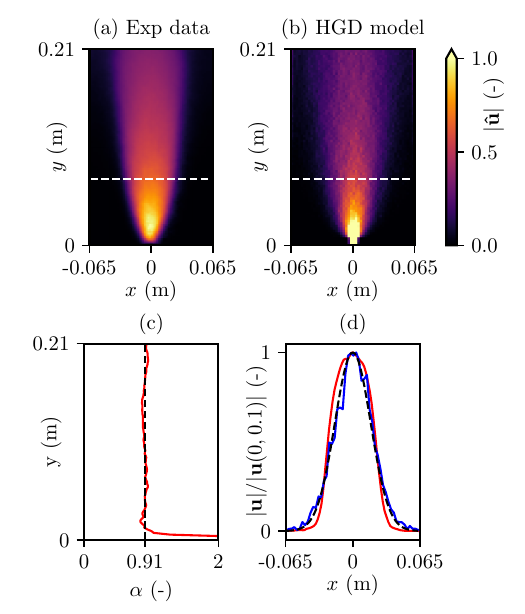}
    \caption{Comparison of silo discharge velocity fields obtained with X-ray radiography in~\cite{guillard2017dynamic} against simulated output with varying $\alpha$. (a): Experimental data from \cite{guillard2017dynamic}. Colour represents normalised velocity magnitude of the discharge of Jasmine Rice through a hopper with a slit opening at the base of width 10 mm. Experimental velocity fields are normalised by the maximum measured value, which occurs above the outlet. Velocities appear to go to zero at the outlet due to limitations of the experimental method. (b) Equivalent simulated velocity field for the same initial conditions. Simulated values are normalised by the value at position (0,0.03). (c) Value of $\alpha$ calculated from the experimental data above via best fit of a Gaussian function at each elevation $y$ with a standard deviation of $\sqrt{2\alpha\bar{s}y}$ and $\bar s=2.3$~mm. (d) Comparison of the experimental (red), simulated (blue) and expected (dashed black, see \ref{app:plume}) normalised velocity fields at an elevation of $y=0.07$~m, indicated with a white dashed line in (a) and (b).}
    \label{fig:silo}
\end{figure}

Finally, we introduce a simple method for the calibration of the material parameter $\alpha$. During the discharge of a silo, we observe a plume of voids which enter the system at the outlet and migrate upward. For a narrow outlet, the standard deviation of the plume width at a height $y$ above the outlet can be shown to be $\sqrt{2\alpha\bar{s}y}$ (see \ref{app:plume}). By measuring the width of this plume, it is then possible to estimate $\alpha$, as shown in Figure~\ref{fig:silo}, where model output is compared with measured velocity fields from X-ray radiography in~\cite{guillard2017dynamic}. Particle image velocimetry of X-ray radiography by \cite{guillard2017dynamic} provides measurements of the plume width for the discharge of Jasmine Rice with nominal size of 2.3~mm. Accordingly, we find a value $\alpha\approx 0.91$ at all elevations except near the outlet where the particle image velocimetry does not give reliable measurements.

\section{Conclusion}

This paper establishes a conceptual framework for a stochastic granular modelling paradigm termed Heterarchical Granular Dynamics (HGD). Positioned as a bridge between particle-level and continuum approaches (see Table~\ref{tab:methods}), HGD combines stochastic void migration, multiscale heterarchical physics, and stress calculations into a unified computational approach. While further validation and refinement are essential, this paper provides the first working example of an HGD model and sets the stage for its application to diverse and complex granular flow scenarios.


Key aspects of the model include its ability to describe both flowing and stable states, segregation and mixing behaviour, and stress distributions in granular materials. The void migration model is validated against x-ray radiography data of silo discharge and synthetic granular column collapse tests. The stress model is validated against basal measurements of the vertical stress.

The model departs from traditional particle-based simulations, enabling the representation of systems with arbitrarily large numbers of grains. However, significant opportunities for further refinement remain. For instance, while the model conserves mass, future developments should incorporate momentum and energy conservation by coupling the stochastic void migration with the force propagation model. Additionally, the absence of features such as inertia, particle breakage, and compaction mechanisms represents areas for future exploration and enhancement.


\section*{Acknowledgements}
The authors would like to thank Fran\c{c}ois Guillard for insightful discussions which helped to formulate this model. 

\section*{Code availability}
All code used to generate the Figures in this paper can be found at \url{https://github.com/benjym/HGD}.

\section*{Declaration of generative AI and AI-assisted technologies in the writing process}
During the preparation of this work the authors used ChatGPT in order to assist with editing. After using this tool/service, the authors reviewed and edited the content as needed and take full responsibility for the content of the publication.

\appendix

\section{Numerical dispersion}\label{app:dispersion}

\subsection{Advection}
Assume the cells move with a constant velocity $u$. For a probability of motion $P$, movement occurs every $T=\tfrac{1}{P}$ timesteps on average, but the actual timestep at which the particle moves is randomly chosen. Let the total number of timesteps be $N_t$. Define a sequence of random variables $T_i$ which denote the timesteps when the particle actually moves. Each $T_i$ is drawn from a distribution where the expected interval is $T$ timesteps. The total displacement $X_N$ of the particle after $N_t$ timesteps is the sum of individual displacements
\begin{equation}
    X_N = u \sum_{i=1}^{q} \Delta t_i,
\end{equation}

\noindent where $q$ is the total number of moves the particle makes and $\Delta t_i$ is the time interval between successive moves. On average, the particle moves every $T$ timesteps, so the number of moves $q$ can be approximated by $\frac{N_t}{T}$. The expected displacement after $N_t$ timesteps is
\begin{equation}
    \mathbb{E}[X_N] = u \sum_{i=1}^{q} \mathbb{E}[\Delta t_i] = u \cdot q \cdot \mathbb{E}[\Delta t] = u \cdot \frac{N_t}{T} \cdot T \Delta t = u N_t \Delta t.
\end{equation}

Given that $t = N_t \Delta t$, the ensemble average after time $t$ is
\begin{equation}
    \mathbb{E}[X(t)] = u t.
\end{equation}

Each move introduces some randomness, but since the moves are perfectly correlated with the velocity $u$, the primary source of variance comes from the intervals $\Delta t_i$. If $\Delta t_i$ follows an exponential distribution with mean $T \Delta t$, the variance of $\Delta t_i$ is $(T \Delta t)^2$. Over $q = \frac{N_t}{T}$ moves, the total variance in position due to these random intervals is
\begin{equation}
\text{Var}(X_N) = u^2 \sum_{i=1}^{q} \text{Var}(\Delta t_i) = u^2 \cdot \frac{N_t}{T} \cdot (T \Delta t)^2 = u^2 \cdot N_t \Delta t^2 T,
\end{equation}
with $t = N_t \Delta t$, the variance in position after time $t$ is
\begin{equation}
\text{Var}(X(t)) = u^2 \cdot t \Delta t T.
\end{equation}

The numerical dispersion (standard deviation) is then
\begin{equation}
\sigma_\mathrm{adv} = \sqrt{\text{Var}(X(t))} = \sqrt{u^2 \cdot t \Delta t T} = u \sqrt{t \Delta t T}.
\end{equation}

\subsection{Diffusion}

Assume that a cell diffuses with a diffusivity $D$. In a diffusion process, the variance in position after time $t$ is given by
\begin{equation}\label{eq:variance_diffusion}
    \text{Var}(X(t)) = 2 D t.
\end{equation}

Due to the method of solving the equations described above, cells move at random intervals, occurring every $T$ timesteps on average. Let $\Delta t_i = T_i \Delta t$ be the time intervals between successive moves, where $T_i$ are random variables representing the number of timesteps between moves. When a particle moves due to diffusion, it undergoes a random displacement $\Delta X_i$ with zero mean and variance
\begin{equation}
    \text{Var}(\Delta X_i) = 2 D \Delta t_i.
\end{equation}

The total displacement after $N_t$ timesteps is
\begin{equation}
    X_N = \sum_{i=1}^{q} \Delta X_i,
\end{equation}
where $q$ is the total number of moves, approximately $\frac{N_t}{T}$. The variance in position after $N_t$ timesteps is the sum of the variances from each move
\begin{align}
    \text{Var}(X_N) &= \sum_{i=1}^{q} \text{Var}(\Delta X_i) = \sum_{i=1}^{q} 2 D \Delta t_i = 2 D \sum_{i=1}^{q} \Delta t_i.
\end{align}

Since the total accumulated time from all moves is
\begin{equation}
    \sum_{i=1}^{q} \Delta t_i = t,
\end{equation}
the variance becomes
\begin{equation}
    \text{Var}(X(t)) = 2 D t.
\end{equation}

This is the same as the true variance in Eq.~(\ref{eq:variance_diffusion}). This shows that the variance due to diffusion accumulates linearly with time $t$, independent of the randomness in the intervals $\Delta t_i$. The randomness in the movement intervals $\Delta t_i$ does not contribute additional variance beyond the intrinsic variance from diffusion itself. Unlike the advection case, where the timing of moves affects the variance in position due to the constant velocity $u$, diffusion inherently accounts for random displacements over time.
\section{Discretisation error of slope stability criteria}\label{app:disc}

Because the slope stability metric requires a threshold on $\nu$, it is sensitive to the discretisation imposed via the choice of the number of microstructural cells, $M$, which leads to a finite set of possible $\nu$ values. We can approximate this discretisation effect by assuming that each $\Delta\nu$ value in (\ref{eq:density_cutoff}) deviates from the closest possible value by $\tfrac{1}{2M}$. When $\Delta x = \Delta y$, the expected offset $\Delta\mu$ can be expressed as
\begin{equation}
    \Delta\mu = \frac{1}{\frac{1}{\Delta\nu_\mathrm{lim} + \frac{2}{M}} - 1} - \frac{1}{\frac{1}{\Delta\nu_\mathrm{lim}} - 1}.
\end{equation}

\begin{figure}
    \centering
    \includegraphics{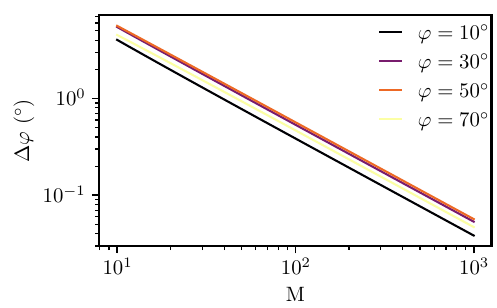}
    \caption{Discretisation error for slope calculation with $\Delta x = \Delta y$.}
    \label{fig:discretisation-error}
\end{figure}

This behaviour is shown in Figure \ref{fig:discretisation-error} for various values of $\Delta\varphi=\tan^{-1}\Delta\mu$. A value of $M\gtrsim60$ ensures $\Delta\varphi < 1^\circ$.

\section{Grainsize dynamics}\label{sec:phi_pde}

To understand the continuum limit of the evolution of the particle sizes, it is necessary to reformulate the conservation equation (\ref{eq:conservation_dense}) in terms of the partial mass of a given particle size. To do this, we first define the solid fraction $\Phi$ of particles between sizes $s_a$ and $s_b$ as
\begin{equation}
\Phi(s_a \le s < s_b) = \frac{1}{\nu M}\underset{\forall s_a\le s < s_b}{\sum^{M}_{k=1}}1.
\end{equation}

As demonstrated in \cite{marks2017heterarchical}, this solid fraction is related to the continuous grainsize distribution $\phi$ by
\begin{equation}
    \Phi(s_a \le s < s_b) = \int_{s_a}^{s_b}\phi(s')~ds'.
\end{equation}



Using these definitions, we can calculate the partial density $\rho = \rho_s\nu\phi$, with $\rho_s$ the density of the solid phase, as
\begin{align}
    \rho_{i,j}^{t+\Delta t} - \rho_{i,j}^t = &+\alpha u s\frac{\Delta t}{\Delta x^2}\rho^t_{i,j}n^t_{i+1,j} \nonumber \\
    &+ \alpha u s\frac{\Delta t}{\Delta x^2}\rho^t_{i,j}n^t_{i-1,j} \nonumber \\
    &- \alpha u s\frac{\Delta t}{\Delta x^2}\rho^t_{i-1,j}n^t_{i,j} \nonumber \\
    &- \alpha u s\frac{\Delta t}{\Delta x^2}\rho^t_{i-1,j}n^t_{i,j} \nonumber \\
    &+ \frac{\overline{s^{-1}}_{i,j}^t}{s} u \frac{\Delta t}{\Delta y}\rho^t_{i,j}n^t_{i,j+1} \nonumber \\
    &- \frac{\overline{s^{-1}}_{i,j-1}^t}{s} u \frac{\Delta t}{\Delta y}\rho^t_{i,j-1}n^t_{i,j}.
\end{align}

This can be rearranged to give
\begin{align}
    \frac{\rho_{i,j}^{t+\Delta t} - \rho_{i,j}^t}{\Delta t} = -\frac{\alpha u s}{\Delta x^2}(
    &-\rho^t_{i,j}n^t_{i+1,j}
    - \rho^t_{i,j}n^t_{i-1,j} \nonumber \\
    &+ \rho^t_{i+1,j}n^t_{i,j}
    + \rho^t_{i-1,j}n^t_{i,j} )\nonumber \\
    + \frac{u}{s\Delta y}&(\overline{s^{-1}}_{i,j}^t\rho^t_{i,j}n^t_{i,j+1} \nonumber\\
    &- \overline{s^{-1}}_{i,j-1}^t\rho^t_{i,j-1}n^t_{i,j}).
\end{align}

By taking the first order terms from a Taylor expansion, these finite differences can be shown to be equivalent to
\begin{equation}
    \frac{\partial\rho}{\partial t} = -n\frac{\partial^2}{\partial x^2}(\alpha u s\rho) + \alpha u s\rho\frac{\partial^2n}{\partial x^2} + \frac{\partial}{\partial y}\left( u\frac{\overline{s^{-1}}}{s}\rho n\right).
\end{equation}


This governing equation contains three terms on the right hand side. The first term represents diffusion of the grainsize due to grainsize concentration gradients. The second term represents size dependent diffusion due to porosity gradients. The final term represents the conventional advective segregation.

Similarly, we can take conservation of mass for the same fraction of particles moving with the barycentric motion, represented by $\bar\rho\bar u$, with $\bar\rho=\int\rho(s')~ds'=\rho_s\nu$, and $\bar u =\int\phi u~ds$, to give
\begin{equation}
    \frac{\partial\bar\rho\phi}{\partial t} = -n\frac{\partial^2}{\partial x^2}(\alpha u \bar{s}\bar\rho\phi) + \alpha u \bar{s}\bar\rho\phi\frac{\partial^2n}{\partial x^2} + \frac{\partial}{\partial y}\left( u\bar\rho\phi n\right).
\end{equation}

Taking the difference between these two expressions, we can define $\hat\rho(s) = \rho(s) - \bar\rho\phi$,
\begin{align}\label{eq:rho_hat}
    \frac{\partial\hat\rho}{\partial t} =
    -\rho_sn\frac{\partial^2}{\partial x^2}(\alpha u\nu\phi(s-\bar{s}))
    + \rho_s\alpha u \nu\phi(s - \bar{s})\frac{\partial^2n}{\partial x^2} \nonumber \\
    +\rho_s\frac{\partial}{\partial y}\left(\left(\frac{\overline{s^{-1}}}{s}-1\right)u\phi\nu n\right).
\end{align}

In this frame of reference, the effects of segregation become clear, with size-dependent processes that go to zero when $s=\bar{s}=\overline{s^{-1}}$, leading to no motion.





\section{Silo discharge void plume width}\label{app:plume}

For particles of a single size, far from boundaries, at the dilute void limit, the model follows Eq.~\ref{eq:dilute_pde}. This advection-diffusion equation can be solved under the assumptions that the medium extends to infinity, and is subject to a steady influx of voids from a point source located at the origin. This boundary value problem is equivalent to the ``Gaussian Plume Model'' used in environmental engineering, where the concentration of a contaminant in a uniform wind (here the contaminant is voids) is calculated \cite{gaussian_plume}. The evolution equation for the porosity field with a source term $S$ of voids can be expressed as
\begin{equation}
    \frac{\partial n}{\partial t} = D \frac{\partial^2 n}{\partial x^2} - u \frac{\partial n}{\partial y} + S(x, y, t).
    \label{eq:advection_diffusion}
\end{equation}

For an instantaneous point source at the origin at time $ t = 0 $
\begin{equation}
    S(x, y, t) = Q \delta(x) \delta(y) \delta(t),
\end{equation}
with $ \delta $ being the Dirac delta function. Due to the advection being in the $y$ direction and diffusion in $ x $, we can write the solution as
\begin{equation}
    n(x, y, t) = \delta(y - ut) \times \frac{Q}{\sqrt{4\pi D t}} \exp\left( -\frac{x^2}{4Dt} \right), \quad t > 0.
    \label{eq:puff_solution}
\end{equation}

The term $ \delta(y - ut) $ accounts for advection in the $y$-direction, indicating that voids move along the line $ y = ut $. The Gaussian term represents diffusion in the $ x $-direction from a point source. For a continuous point source emitting at a constant rate $ Q $, we integrate the instantaneous solutions over time
\begin{equation}
    n(x, y) = \int_{0}^{\infty} Q \delta(y - ut) \frac{1}{\sqrt{4\pi D t}} \exp\left( -\frac{x^2}{4Dt} \right) dt.
    \label{eq:continuous_source}
\end{equation}

Using the property of the Dirac delta function:
\begin{equation}
\delta(y - ut) = \frac{1}{u} \delta\left( t - \frac{y}{u} \right),
\end{equation}
we substitute $ t =\frac{y}{u} $. Since $ t > 0 $, this implies $ y > 0 $ (downstream positions). The integral becomes
\begin{equation}
    n(x, y) = \frac{Q}{\sqrt{4\pi Dyu}} \exp\left( -\frac{x^2 u}{4Dy} \right).
    \label{eq:integrated_solution}
\end{equation}

The standard deviation $ \sigma_x(y) $ of the plume width can then be defined as
\begin{equation}
    \sigma_x(y) = \sqrt{\frac{2Dy}{u}}.
    \label{eq:standard_deviation}
\end{equation}

\bibliographystyle{spphys}       
\bibliography{bib}   

\begin{thebibliography}{10}
\providecommand{\url}[1]{{#1}}
\providecommand{\urlprefix}{URL }
\expandafter\ifx\csname urlstyle\endcsname\relax
  \providecommand{\doi}[1]{DOI \discretionary{}{}{}#1}\else
  \providecommand{\doi}{DOI \discretionary{}{}{}\begingroup
  \urlstyle{rm}\Url}\fi

\bibitem{cundall1979discrete}
P.A. Cundall, O.D. Strack, A discrete numerical model for granular assemblies,
  Géotechnique \textbf{29}(1), 47 (1979)

\bibitem{shekhawat2016toughness}
A.~Shekhawat, R.O. Ritchie, Toughness and strength of nanocrystalline graphene,
  Nature Communications \textbf{7}(1), 10546 (2016)

\bibitem{litwiniszyn1958statistical}
J.~Litwiniszyn, Statistical methods in the mechanics of granular bodies,
  Rheologica Acta \textbf{1}(2), 146 (1958)

\bibitem{mullins1972stochastic}
W.W. Mullins, {Stochastic Theory of Particle Flow under Gravity}, Journal of
  Applied Physics \textbf{43}(2), 665 (1972).
\newblock \doi{10.1063/1.1661175}.
\newblock \urlprefix\url{https://doi.org/10.1063/1.1661175}

\bibitem{bourdeau1989stochastic}
P.L. Bourdeau, M.E. Harr, Stochastic theory of settlement of loose cohesionless
  soils, Géotechnique \textbf{39}(4), 641 (1989).
\newblock \doi{10.1680/geot.1989.39.4.641}

\bibitem{kamrin2007stochastic}
K.~Kamrin, M.Z. Bazant, Stochastic flow rule for granular materials, Physical
  Review E \textbf{75}(4), 041301 (2007)

\bibitem{kamrin2012nonlocal}
K.~Kamrin, G.~Koval, Nonlocal constitutive relation for steady granular flow,
  Physical Review Letters \textbf{108}, 178301 (2012)

\bibitem{dunatunga_kamrin_2015}
S.~Dunatunga, K.~Kamrin, Continuum modelling and simulation of granular flows
  through their many phases, Journal of Fluid Mechanics \textbf{779}, 483–513
  (2015).
\newblock \doi{10.1017/jfm.2015.383}

\bibitem{marks2011cellular}
B.~Marks, I.~Einav, A cellular automaton for segregation during granular
  avalanches, Granular Matter \textbf{13}(3), 211 (2011)

\bibitem{marks2015mixture}
B.~Marks, I.~Einav, A mixture of crushing and segregation: The complexity of
  grainsize in natural granular flows, Geophysical Research Letters
  \textbf{42}(2), 274 (2015)

\bibitem{marks2017heterarchical}
B.~Marks, I.~Einav, A heterarchical multiscale model for granular materials
  with evolving grainsize distribution, Granular Matter \textbf{19}(3), 61
  (2017)

\bibitem{bisht2024heterarchical1}
M.S. Bisht, F.~Guillard, P.~Shelley, B.~Marks, I.~Einav, Heterarchical
  modelling of comminution for rotary mills: part i—particle crushing along
  streamlines, Granular Matter \textbf{26}(4), 88 (2024)

\bibitem{bisht2024heterarchical2}
M.S. Bisht, F.~Guillard, P.~Shelley, B.~Marks, I.~Einav, Heterarchical
  modelling of comminution for rotary mills: part ii—particle crushing with
  segregation and mixing, Granular Matter \textbf{26}(4), 87 (2024)

\bibitem{thompson2007granular}
E.L. Thompson, H.E. Huppert, Granular column collapses: further experimental
  results, Journal of Fluid Mechanics \textbf{575}, 177 (2007)

\bibitem{degaetano2013influence}
M.~Degaetano, L.~Lacaze, J.C. Phillips, The influence of localised size
  reorganisation on short-duration bidispersed granular flows, The European
  Physical Journal E \textbf{36}, 1 (2013)

\bibitem{lai2017collapse}
Z.~Lai, L.E. Vallejo, W.~Zhou, G.~Ma, J.M. Espitia, B.~Caicedo, X.~Chang,
  Collapse of granular columns with fractal particle size distribution:
  Implications for understanding the role of small particles in granular flows,
  Geophysical Research Letters \textbf{44}(24), 12 (2017)

\bibitem{he2021experimental}
K.~He, H.~Shi, X.~Yu, An experimental study on aquatic collapses of bidisperse
  granular deposits, Physics of Fluids \textbf{33}(10) (2021)

\bibitem{harr1977mechanics}
M.~Harr, \emph{Mechanics of Particulate Media: A Probabilistic Approach}
  (McGraw-Hill, New York, 1977)

\bibitem{nedderman1979kinematic}
R.~Nedderman, U.~T{\"u}z{\"u}n, A kinematic model for the flow of granular
  materials, Powder Technology \textbf{22}(2), 243 (1979)

\bibitem{choi2005velocity}
J.~Choi, A.~Kudrolli, M.Z. Bazant, Velocity profile of granular flows inside
  silos and hoppers, Journal of Physics: Condensed Matter \textbf{17}(24),
  S2533 (2005)

\bibitem{cooke1979interparticle}
M.H. Cooke, J.~Bridgewater, Interparticle percolation: a statistical mechanical
  interpretation, Industrial \& Engineering Chemistry Fundamentals
  \textbf{18}(1), 25 (1979)

\bibitem{hrenikoff1941solution}
A.~Hrennikoff, Solution of problems of elasticity by the framework method,
  Journal of Applied Mechanics \textbf{8}(4), A169 (1941)

\bibitem{claudin1998models}
P.~Claudin, J.P. Bouchaud, M.~Cates, J.~Wittmer, Models of stress fluctuations
  in granular media, Physical Review E \textbf{57}(4), 4441 (1998)

\bibitem{marks2017size}
B.~Marks, J.A. Eriksen, G.~Dumazer, B.~Sandnes, K.J. M{\aa}l{\o}y, Size
  segregation of intruders in perpetual granular avalanches, Journal of Fluid
  Mechanics \textbf{825}, 502 (2017)

\bibitem{guillard2016scaling}
F.~Guillard, Y.~Forterre, O.~Pouliquen, Scaling laws for segregation forces in
  dense sheared granular flows, Journal of Fluid Mechanics \textbf{807}, R1
  (2016).
\newblock \doi{10.1017/jfm.2016.605}

\bibitem{trollope1956stability}
D.H. Trollope, The stability of wedges of granular materials.
\newblock Ph.D. thesis, University of Melbourne, School of Engineering (1956)

\bibitem{guillard2017dynamic}
F.~Guillard, B.~Marks, I.~Einav, Dynamic x-ray radiography reveals particle
  size and shape orientation fields during granular flow, Scientific Reports
  \textbf{7}(1), 8155 (2017)

\bibitem{gaussian_plume}
W.J. Veigele, J.H. Head, Derivation of the gaussian plume model, Journal of the
  Air Pollution Control Association \textbf{28}(11), 1139 (1978).
\newblock \doi{10.1080/00022470.1978.10470720}

\end{thebibliography}

\end{document}